\documentclass[%
 reprint,
 amsmath,amssymb,
 aps,linenumbers,
]{revtex4-1}

\usepackage{graphicx} 
\usepackage{dcolumn}  
\usepackage{bm}       
\usepackage{verbatim} 
\usepackage{gensymb}
\usepackage{lineno}   
\usepackage{soul}     %
\usepackage{color}
\usepackage{textcomp}

\begin{document}
\nolinenumbers

\preprint{APS/123-QED}

\title{Three-dimensional trapping and assembly of small particles with synchronized spherical acoustical vortices}

\author{Zhixiong Gong}
\affiliation{Univ. Lille, CNRS, Centrale Lille, Yncréa ISEN, Univ. Polytechnique Hauts-de-France, UMR 8520 - IEMN, F- 59000 Lille, France.}%
\author{Michael Baudoin}%
\email{Corresponding author: michael.baudoin@univ-lille.fr}
\homepage{\mbox{http://films-lab.univ-lille1.fr/michael}}
\affiliation{Univ. Lille, CNRS, Centrale Lille, Yncréa ISEN, Univ. Polytechnique Hauts-de-France, UMR 8520 - IEMN, F- 59000 Lille, France.}%
\affiliation{Institut Universitaire de France, 1 rue Descartes, 75005 Paris}%

\date{\today}
             
\begin{abstract}
Three-dimensional harmless contactless manipulation and assembly of micro-objects and micro-organisms would open new horizons in  microrobotics and microbiology, e.g. for microsystems assembly or tissue engineering. In our previous work [Gong \& Baudoin, Phys. Rev. Appl., 12: 024045 (2019)], we investigated theoretically the possibility to trap and assemble in two dimensions small particles compared to the wavelength with synchronized acoustical tweezers based on cylindrical acoustical vortices. However, since these wavefields are progressive along their central axis, they can only push or pull (not trap) particles in this direction and hence are mainly limited to 2D operations. In this paper, we extend our previous analysis and show theoretically that particles can be trapped and assembled in three-dimensions with synchronized spherical vortices. We show that the particles can be approached both laterally and axially and we determine the maximum assembly speed by balancing  the Stokes' drag force and the critical radiation force. These theoretical results provide guidelines to design selective acoustical tweezers able to trap and assemble particles in three dimensions.
\end{abstract}

\pacs{Valid PACS appear here}
\maketitle

\section{\label{sec:Introduction}Introduction}

The father of single beam optical tweezers, Arthur Ashkin, was awarded the Nobel prize in physics in 2018 for his contribution to "groundbreaking inventions in the field of laser physics", which opened the doors to major breakthroughs in physics including neutral atom trapping and cooling \cite{steven1998manipulation}, but also the manipulation of objects ranging from atoms \cite{chu1986experimental} to Bose-Einstein condensates \cite{anderson1995observation} or living cells and bacteria \cite{nat_ashkin_1987}. Yet, optical tweezers have stringent limitations in life science,  which are prohibitive for many applications: (i) The forces which can be applied with optical tweezers on microorganisms are limited to typically less than $50$ pN \cite{m_keloth_2018} and require high intensity beams. This is due to the fact that the optical radiation force used to trap objects is proportional to the intensity of the incident field divided by the, high value, light speed.  (ii) These high intensity beams induce phototoxicity on biosamples due to thermal effects and/or chemical reactions \cite{bj_liu_1995,bj_liu_1996,Neuman1999,m_keloth_2018,Blazquez2019}. Finally (iii) objects can only be manipulated in optically transparent media, that prevent their use for most in-vivo application. The remote contactless manipulation of micro-objects can also be achieved with magnetic tweezers \cite{ecr_crick_1950,s_strick_1996,s_strick_1996}. These latter are biocompatible and easy to implement but (i) they can only be used to move magnetic objects \cite{schuerle2019synthetic} or otherwise require pre-tagging and (ii) they exhibit low trap stiffness and low selectivity. 

Many limitations of optical and magnetic tweezers for microbiology applications can be overcome with acoustical tweezers \cite{baudoin2020Acoustical}: (i) Since the trapping force applied with acoustical tweezers is, as their optical counterpart, proportional to the incident field intensity divided by the wave speed, the drastically lower speed of sound compared to light leads to trap strengths some orders of magnitude larger than with optical tweezers at same driving intensity \cite{baresch2016observation,nc_baudoin_2020}. (ii) Acoustical tweezers can trap a large variety of materials since only a contrast in density and/or compressibility between the particle and the surrounding medium is required for the acoustic radiation force to exist. (iii) Since acoustic transducer are available from kHz to GHz frequencies in liquids, the manipulation of particles with sizes ranging from centimeter to nanometer sizes can be envisioned. (iv) Acoustical waves are highly biocompatible for both in-vivo \cite{ap_szabo_2014} and in-vitro \cite{umb_stuart_2000,umb_humstrom_2007,wiklund2012,po_burguillos_2013,marx2015} applications.

The concept of acoustical tweezers was first introduced by Wu et al. in 1991  \cite{wu1991acoustical}, in analogy with optics. But in this early work, axial trapping was achieved with two counter-propagating beams, hence requiring to position transducers on both sides of the trapping area. Three-dimensional trapping \cite{baresch2016observation}, levitation \cite{marzo2015holographic} and selective manipulation of microparticles \cite{baudoin2019folding} and cells \cite{nc_baudoin_2020} with single beam acoustical tweezers (i.e. with a wavefield synthesis system located on only one side of the manipulation area) has been achieved only recently by using specific wavefields called focused (spherical) acoustical vortices. Indeed, classical focused waves cannot be used as in optics since many particles of interest such as cells or solid particles would be expelled from the focal point. One-sided focused acoustical vortices on the other hand provide a 3D trap for such particles as first demonstrated by Baresch et al. \cite{baresch2013spherical}. They also enable strong spatial localization of the wavefield, that is  required to be selective, i.e. to be able to pick up  and move a single particle independently of its neighbors. Beyond the manipulation of individual particles, one key operation in microfluidics and microbiology is to assemble objects. This feature is essential e.g. to assemble spheroids or for tissue engineering.

A first strategy to move and assemble multiple objects is to adapt dynamically the acoustic field to the target task, as demonstrated in 2D by Courtney et al. \cite{courtney2014}. Such adaptive field synthesis strategy nevertheless requires complex array of transducers that (i) are only presently available at  low frequency for relatively large particle trapping and (ii) are opaque, cumbersome and hence not compatible with classical microscopy environments. In addition, while advanced 3D manipulation of multiple objects has been demonstrated with two arrays of transducers located on both side of the manipulation area \cite{pnas_marzo_2019}, the possibility to trap, move and assemble in 3D multiple objects with single beam acoustical tweezers  has not been demonstrated yet.  A second strategy is to trap each particle to be assembled at the core of an acoustical vortex and use the interference between synchronized acoustical vortices to assemble them. This possibility has been investigated for the 2D assembly of small particles compared to the wavelength with cylindrical acoustical vortices by Gong \& Baudoin  \cite{prap_gong_2019}. The advantage of this solution is that it can be implemented for both reconfigurable arrays of transducers \cite{hefner1999,prl_marchiano_2003,prl_volke_2008,njp_skeldon_2008,courtney2014,marzo2015holographic,prap_riaud_2015,ieee_riaud_2016} and static holographic vortices wave synthesis systems \cite{hefner1999,jasa_gspan_2004,ieee_elao_2011,pre_jimenez_2016,prl_jiang_2016,apl_jiang_2016,apl_naify_2016,apl_wang_2016,prap_riaud_2017,mupb_terzi_2017,apl_jimenez_2018,apl_muelas_2018,baudoin2019folding,sr_jimenez_2019,nc_baudoin_2020}. In this paper, we extend our previous work to the three dimensional case: we demonstrate theoretically that particle can be trapped, moved and assembled in 3D by using spherical (focused) acoustical vortices of first topological order. While further numerical and experimental investigation is necessary to investigate the case of vortices synthesized with a finite aperture, this work prefigures the use of focused acoustical vortices for assembly of particles with single beam acoustical tweezers.

\section{\label{sec: Single SBB}3D trapping in a single spherical Bessel acoustical vortex}

\subsection{Spherical Bessel acoustical vortices}

\begin{figure} [!htbp]
\includegraphics[width=8.6cm]{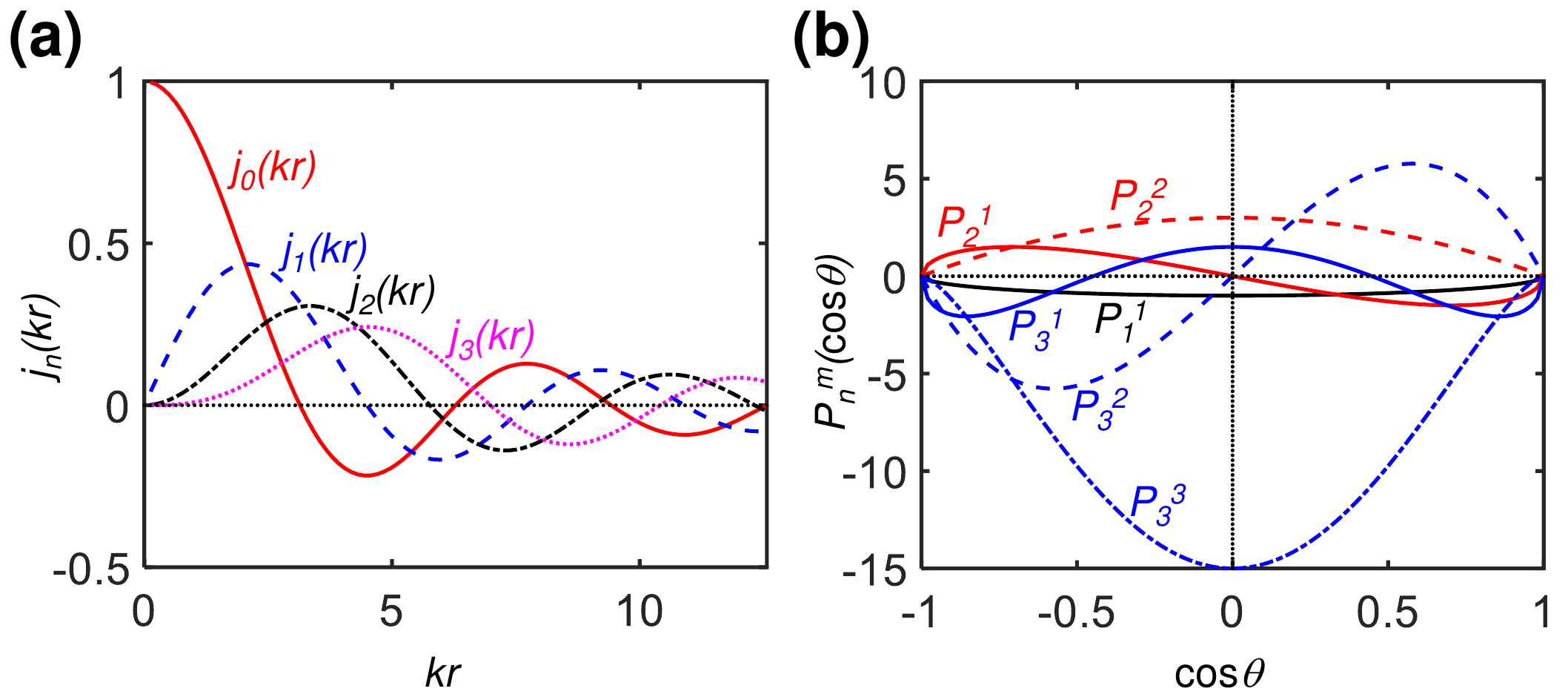}
\caption{(a) Spherical Bessel functions $j_n(kr)$ magnitude as a function of the dimensionless radial distance $kr$ with $k = \omega / c_0$ the wavenumber. (b) Associated Legendre functions magnitude $P_n^m(\cos\theta)$ as a function of $\cos\theta \in [-1,1]$.}
\label{Figure1}
\end{figure}

Cylindrical and spherical bessel acoustical vortices are separate variable solutions of d'Alembert wave equation in cylindrical and spherical coordinates respectively \cite{baudoin2020Acoustical}. These waves are spinning around a phase singularity wherein the amplitude cancels, surrounded by a ring of high intensity. Cylindrical vortices, first investigated in acoustics by Hefner \& Marston \cite{hefner1999} are laterally focused waves. These waves are interesting to trap particles laterally (see e.g. \cite{prap_riaud_2017}) but they can only push or pull  (and not trap) particles along their propagation axis depending on the particles and beam properties \cite{fan2019trapping}. Spherical acoustical vortices on the other hand focus the energy in three dimensions and hence can create a 3D trap even with a finite aperture \cite{baresch2013spherical}. Spherical Bessel beams are defined by the equation \cite{baudoin2020Acoustical}: 
\begin{equation}
p=A_{0} j_{n}\left(k r\right) P_n^m (\cos\theta) e^{i (m \varphi - \omega t)},
\label{eq:sbb}
\end{equation}
with $A_0$ the beam amplitude,  $j_n(kr)$ the spherical Bessel function of the first kind of order $n$, $k = \omega / c_0$ the wavenumber, $\omega$ the angular frequency, $c_0$ the wave speed, $r$ the radius, $P_n^m(\cos\theta)$  the associated Legendre polynomials of order ($n,m$), $\theta$ and $\varphi$ are the polar and azimuthal angles in spherical coordinates, $z$ is the central axis of the vortex, $(x,y)$ the lateral directions and $t$ the time. Finally $m$ is the topological charge of the spherical Bessel beams verifying $-n \leq m \leq n$, which describes the periodic number of phase change from 0 to $2 \pi$ in the transverse plane. The Bessel functions and associated Legendre polynomials are represented on Fig. \ref{Figure1}a and \ref{Figure1}b respectively . Spherical Bessel acoustical vortices correspond to Bessel beams of order $m\geq 1$, since in this case the phase is spinning around the $z$-axis. The case $m=0$ however, corresponds to spherically focused waves. This case will not be further discussed in this paper since particles with positive contrast factors with respect to the surrounding fluids (like solid particle surrounded by a fluid) are expelled (not trapped) from the center of the beam in the long wavelength regime (LWR) $ka \ll 1$. Finally, we can note that the average power $\left< \mathcal{P} \right>$ required to synthesize an acoustical vortex of amplitude $A_0$ is  proportional to (see Appendix \ref{sec:Appendix A}):
\begin{equation}
\mathcal{\left< \mathcal{P} \right>} \propto \frac{2 \pi A_0^2}{\rho_0 \omega} \times \frac{2 (n+m)!}{(2n+1) (n-m)!}.
\label{power}
\end{equation} 
Hence to compare the trapping capabilities of acoustic beams of different orders $(n,m)$ at same input power, we will normalize the amplitude $A_0$ by $\sqrt{(2 n+1)(n-m)!} / \sqrt{2(n+m)!}$ in the next section.

\subsection{Gradient and scattering forces in the long wavelength approximation}

\begin{figure*} [!htbp]
\includegraphics[width=17.8cm]{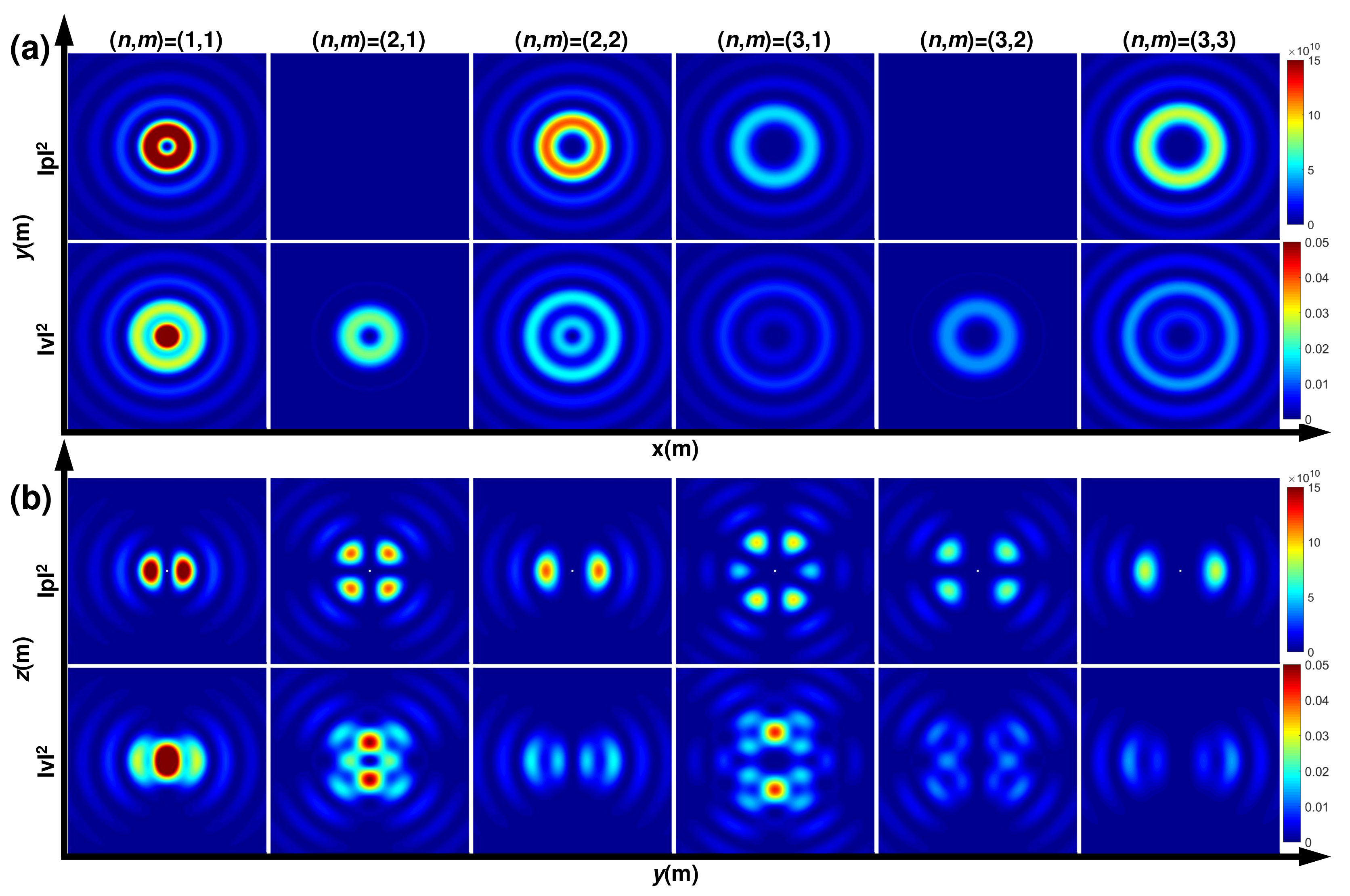}
\caption{Square of the pressure modulus $|p|^{2}$  and square of the velocity modulus $|\mathbf{v}|^{2}$ in the ($x,y$) plane (subfigure (a)) and in the ($y,z$) plane (subfigure (b)) for a single spherical Bessel vortex (SBV) with different orders ($n,m$) and the same input power (corresponding to an amplitude $A_0=10^6$ for ($n,m$)=(1,1) and normalized according to Eq. \ref{power} for other orders).}
\label{Figure2}
\end{figure*}

The force applied on particles much smaller than the wavelength ($ka \ll 1$) can be calculated with the general formula \cite{baresch2016observation}:
\begin{align}
\mathbf{F} = & - V_0 \left\{ \nabla \left[ f_1 \left( \frac{|p|^2}{4 \rho_0 c_0^2} \right) - f_2 \left( \frac{\rho_0 | \mathbf{v} |^2}{4} \right) \right]  \right. \label{eq:gorkov} \\
& - \frac{(ka)^3}{3} \left[  \left( f_1^2 + \frac{f_1 f_2}{3} \right)  
\mathrm{Re} \left( \frac{k}{2 c_0} p \mathbf{v}^* \right) \right.  \nonumber \\
&   \left. + \left. \frac{f_2^2}{3} \mathrm{Im} \left( \frac{\rho_0}{2} \mathbf{v}. \nabla \mathbf{v}^* \right)  \right]  \right\} \nonumber 
\end{align}
with $V_0 = 4/3 \, \pi a^3$ the volume of the particle, $a$ its radius, $f_1 = (1 - \kappa_p / \kappa_0)$ the monopolar acoustic contrast factor related to the particle isotropic expansion/compression, $\kappa_p$ and $\kappa_0$ the compressibility of the particle and fluid respectively, $f_2 = 3 (\rho_p - \rho_0) / (2 \rho_p + \rho_0)$ the dipolar acoustic contrast factor related to the particle back and forth translation, $\rho_p$ and $\rho_0$ the density of the particle and fluid respectively, $p$ and $\mathbf{v}$ the complex pressure and velocity of the incident acoustic fields, and finally ``Re" and ``Im'' designate respectively the real and imaginary part of a complex number and the supercript ``$^*$" stands for the complex conjugate.

This expression can be divided into two main types of contributions: the gradient force $\mathbf{F_{grad}}$ and the scattering force $\mathbf{F_{scat}}$. The force $\mathbf{F_{grad}}$ constituted by the terms on the rhs of the first line of Eq. (\ref{eq:gorkov}) can be expressed as the negative gradient of a potential $U$, known as the Gor'kov's potential \cite{Gorkov1962on}:
\begin{align}
& \mathbf{F_{grad}} = - \nabla U \\
& \mbox{with } U = V_0  \left[ f_1 \left( \frac{|p|^2}{4 \rho_0 c_0^2} \right) - f_2 \left( \frac{\rho_0 | \mathbf{v} |^2}{4} \right) \right]. \label{gorkov}
\end{align}
This gradient force, as its name suggests, results from gradients of the acoustic field magnitude. The first term is a monopolar contribution related to the potential energy density of the acoustic field $\mathcal{V} = |p|^2 / 4 \rho_0 c_0^2$, while the  second term is a dipolar contribution related to the kinetic energy density of the acoustic field $\mathcal{K} = \rho_0 | \mathbf{v} |^2 / 4$ \footnote{N.B. The factor 4 instead of 2 in the potential and kinetic energies comes from the fact that we consider the square of the modulus of the complex expressions of the pressure and velocity, that are equal to 2 times the time average of the square of the real expressions of the pressure and velocity fields}. We can see from this expression that when the monopolar acoustic constrast factor $f_1$ is positive (i.e. for particles less compressible than the surrounding fluid), particles are pushed by the potential energy toward the nodes (minima) of the pressure field. While when the dipolar acoustic contrast factor $f_2$ is positive (i.e. for particles more dense than the surrounding fluid) particles are pushed toward the antinodes (maxima) of the velocity field. Of course for plane standing wave, pressure nodes correspond to velocity antinodes, so that particles more dense and less compressible than the surrounding fluid are pushed by both the potential and kinetic energy toward the same location. We will see in the next section that depending on their order $(n,m)$, things are not so obvious for acoustical vortices.  

The terms expressed in the second and third line of Eq. (\ref{eq:gorkov}) are known as the scattering force $\mathbf{F_{scat}}$. It contains three terms: one due to the monopolar oscillation of the particle only ($\propto f_1^2$), one due to dipolar oscillation only ($\propto f_2^2$) and one due to cross coupling between monopolar and dipolar oscillations $(\propto f_1 f_2)$.

It is interesting to note that for an incident standing wave, only the gradient force acts on the particle, while the scattering force cancels. On the opposite, for a progressive wave, only the scattering force acts on the particle and the gradient force cancels. This can be seen directly from Eq. (\ref{eq:gorkov}). Note that this analysis is only valid in the long wavelength regime (LWR) $ka \ll 1$. Another interesting point is that the scattering forces are proportional to a factor $(ka)^3$, which is very small in the LWR. Hence generally speaking, scattering forces are weak compared to gradient forces in this regime. As can be seen from Eq. (\ref{eq:sbb}) the SBV is neither a standing nor a progressive wave. It is a standing wave over $r$ and $\theta$, while it is progressive over $\phi$. Hence we expect a gradient force over $\mathbf{e}_r$ and $\mathbf{e_\theta}$ and a scattering force over $\mathbf{e_\varphi}$, with $(\mathbf{e}_r,\mathbf{e_\theta},\mathbf{e_\varphi})$ the spherical basis. To verify this hypothesis, we can compute the gradient and scattering forces. For the gradient force, we can start by calculating the square of the modulus of the complex pressure and velocity. We obtain:
\begin{equation}
|p|^2 = pp^* = A_0^2 j_n^2 (kr) {P_n^m }^2(\cos\theta)
\label{eq:p2}
\end{equation}
and since:
\begin{align}
\mathbf{v} = & - \frac{i}{\rho_0 \omega} \nabla p \nonumber \\
  = &  - \frac{i A_0}{\rho_0 \omega} e^{i(m \varphi - \omega t)} \left[  k j_n'(kr) P_n^m(\cos \theta) \mathbf{e}_r \right. \label{eq:v}   \\
  & + \left. \frac{j_n(kr)}{r} \frac{d P_n^m(\cos \theta)}{d \theta} \mathbf{e_\theta} + im \frac{j_n(kr)}{r} \frac{P_n^m(\cos \theta )}{\sin \theta} \mathbf{e_\varphi} \right] \nonumber
\end{align}
with $i$ the imaginary unit, we obtain:
\begin{align}
|\mathbf{v}|^2 & = \frac{ A_0^2}{\rho_0^2 \omega^2} \left[  k^2 (j_n'(kr))^2 {P_n^m}^2(\cos \theta)   \right. \label{eq:v2} \\
  & + \left. \frac{{j_n}^2(kr)}{r^2}  \left( \frac{d P_n^m(\cos \theta)}{d \theta} \right)^2 + m^2 \frac{{j_n}^2(kr)}{r^2} \frac{{P_n^m}^2(\cos \theta )}{\sin^2 \theta} \right] \nonumber
\end{align}
The variation of $|p|^2$ and $|\mathbf{v}|^2$ in the $(x,y)$ and $(y,z)$ planes are represented on Fig. \ref{Figure2}a and \ref{Figure2}b respectively for different orders (n,m). Movie 1 in SI also shows 3D views of the pressure amplitudes square of SBVs with different orders for better representation of the entire beam structure. Since the gradient force $\mathbf{F_{grad}}$ is proportional to the gradients of $|p|^2$ and $|\mathbf{v}|^2$, which depend only on $r$ and $\theta$ according to Eqs. (\ref{eq:p2}) and (\ref{eq:v2}), we see, as expected, that this force has only some components along $\mathbf{e}_r$ and $\mathbf{e_\theta}$, along which the wave is a standing wave.

Now, we can look at the scattering force. Straightforward calculation gives:
\begin{equation}
\mathrm{Re}(p \mathbf{v}^*) = \frac{m A_0^2}{\rho_0 \omega} \left[ \frac{{j_n}^2(kr)}{r} \frac{{P_n^m}^2(\cos \theta )}{\sin \theta} \right] \mathbf{e_\varphi}
\end{equation}
and:
\begin{align}
\mathrm{Im}(\mathbf{v}. \nabla \mathbf{v}^*) & = \left[ v_r \frac{\partial v_\varphi^*}{\partial r} + \frac{1}{r} \left( v_\theta \frac{\partial v_\varphi^*}{\partial \theta} + \frac{v_\varphi}{\sin \theta} \frac{\partial v_\varphi^*}{\partial \varphi} \right. \right. \\
& + \left. \left. v_r v_\varphi^* + \frac{v_r v_\varphi^*}{\tan \theta} \right) \right]\mathbf{e_\varphi} \nonumber
\end{align}
where $(v_r, \, v_\theta, \, v_\varphi)$ are the components of the complex acoustic velocity over $(\mathbf{e}_r, \mathbf{e_\theta}, \mathbf{e_\varphi})$ respectively given by Eq. (\ref{eq:v}). The scattering force is, as expected, directed toward the only direction wherein the wave is progressive, i.e. $\mathbf{e_\varphi}$. Since this scattering force is small compared to the gradient force as discussed previously, we will neglect it in the subsequent analysis.

\begin{figure} [!htbp]
\includegraphics[width=8.6cm]{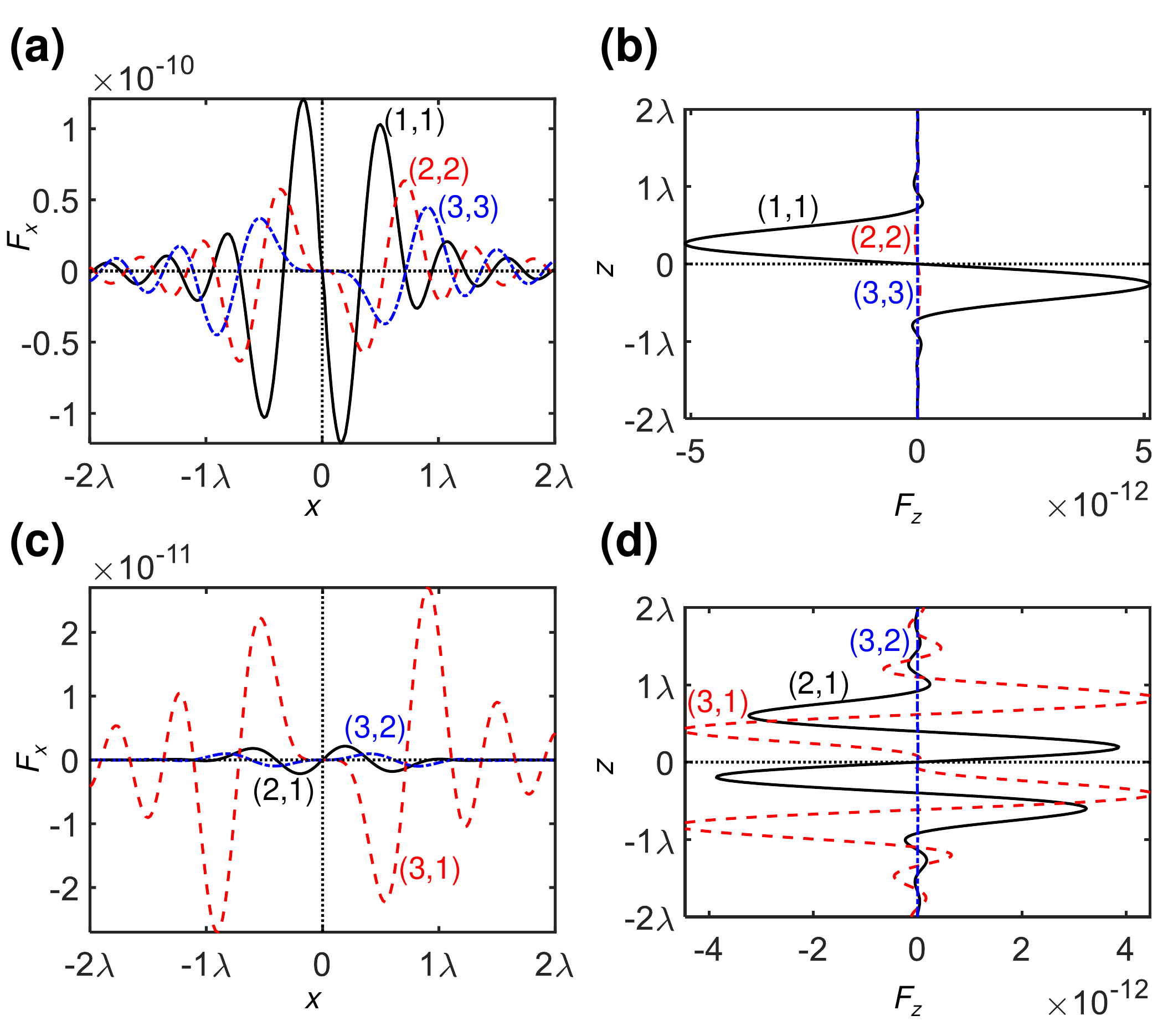}
\caption{Gradient lateral force $F_x$ and axial force $F_z$ exerted on an elastic polystyrene (PS) particle of radius $a=5 \; \mu$m immersed in water by a spherical Bessel vortex of order $(n,m)$ at driving frequency $5$ MHz. (a) and (c) Lateral forces for the orders $(n=m)$ and $(n \neq m)$ respectively. (b) and (d) Axial forces for the orders $(n=m)$ and $(n \neq m)$, respectively.}
\label{Figure3}
\end{figure}

\subsection{Lateral force ($(x,y)$ plane)}

\begin{table*}[!htbp]
\small
  \caption{ Acoustic properties}
  \label{table1}
  \begin{tabular*}{1\textwidth}{@{\extracolsep{\fill}}lcccc}
    \hline 
    Material & Density $\rho_0$ ($kg/m^3$) & Compressibility $\kappa$ (1/TPa) & Longitud. speed of sound $c$ (m/s) & Viscosity $\eta$ (mPa s)\\
    \hline 
    PS       & 1050   & 172   & 2350   & ...          \\
    Water    & 998.2  & 456   & 1482   & 1.002        \\
    \hline
  \end{tabular*}
\end{table*}

In this subsection we analyse the ability of a SBV to trap laterally a particle with positive contrast factors $f_1$ and $f_2$  at the center of the vortex (i.e in the $(x,y)$ plane at the altitude $z=0$) depending on the order of the vortex $(n,m)$ (with $m >0$). The $(x,y)$ plane corresponds to $\theta = \pi/2$, i.e. $\cos(\theta) =0$. Since $P_n^m(-x) = (-1)^{(n+m)} P_n^m (x)$, if $(n+m)$ is odd, then $P_n^m$ is a odd function and $P_n^m(0) = 0$. \\

\textbf{Case A: (n+m) is odd.} In this case, since $|p|^2 \propto {P_n^m}^2(\cos\theta)$ (see Eq. (\ref{eq:p2})), the potential energy is null. This can be seen on Fig. \ref{Figure2}a for $(n,m) = (2,1) \mbox{ and } (3,2)$. Hence the potential energy does not contribute to lateral forces. On the other hand, only the term $(d P_n^m(\cos\theta) / d \theta)^2$ does not vanish in the expression of the kinetic energy (Eq. (\ref{eq:v2})). If we introduce the classical relation:
\begin{align}
& \frac{d}{d \theta} P_n^m (\cos \theta) \\
& = \frac{1}{\sin(\theta)} \left[n \cos \theta P_n^m (\cos \theta) - (n+m) P_{n-1}^m (\cos \theta)) \right], \nonumber
\end{align}
we see that when $(n+m)$ is odd, only the term $P_{n-1}^m(\cos(\theta))$ does not vanish when $\theta = \pi/2$. Thus the potential becomes equal to:
$$
U = - \frac{V_0 f_2 A_0^2}{4 \rho_0 \omega^2}  \frac{{j_n(kr)}^2}{r^2} (n+m)^2 \frac{{P_{n-1}^m}^2(\cos \theta)}{\sin^2 \theta}
$$
In this expression the lateral variation is given by the function $j_n(kr)^2/r^2$ with $n \geq 2$, which becomes minimum (null) when $r \rightarrow 0$ (see Fig. \ref{Figure1}). Hence since $U$ is negative for particles with positive acoustic contrast factor $f_2$ (i.e. particle more dense that the surrounding fluid), and $\mathbf{F_{grad}} = - \nabla U$, such particles are expelled from the center of the vortex when $(n+m)$ is odd. This can indeed be seen on Fig. \ref{Figure3}c representing the lateral force $F_x$ for an elastic  polystyrene (PS) particle of radius $a=5 \; \mu$m immersed in water and insonified with a SBV at the driving frequency $5$ MHz. The properties used for the simulation are summarized in table \ref{table1} and are used for all the simulations presented in this paper. \\

\textbf{Case B: (n+m) is even.} When $(n+m)$ is even, we can distinguish two cases. For $(n,m) = (1,1)$ both the potential energy and kinetic energy contribute to trap particles with positive contrast factors $f_1$ and $f_2$ at the vortex center. Indeed, the potential energy is minimum at the center, while the kinetic energy is maximum (see  Fig. \ref{Figure2}). For the cases $(n,m) \neq (1,1)$ we can see on Fig. \ref{Figure2} that both the potential and kinetic energy are minimum at the center of the vortex. Hence the potential energy contributes to the lateral trap, while the kinetic energy tends to expel particles with positive contrast factor $f_2$ from the center. Hence depending whether the compressibility or density contrast is larger the particle can be either trapped or expelled.

\subsection{Axial force (z-direction)}

In this subsection we analyse the ability of a SBV to trap axially (along the $z$-axis) a particle with positive contrast factors $f_1$ and $f_2$ at the center of the vortex, depending on the order of the vortex $(n,m)$ (with $m >0$). The $z$-axis corresponds to $\theta = \{0, \, \pi \}$, i.e., $\cos(\theta) = \pm 1$. Due to the phase singularity on this axis, $P_n^m(\pm1) = 0$ $\forall m \neq 0$ (see Fig. \ref{Figure1}b). Hence from Eqs. (\ref{eq:p2}) and (\ref{eq:v2}), we see that for $m \geq 1$:
\begin{align}
& |p|^2 = 0, 
& |\mathbf{v}|^2 = \frac{ A_0^2}{\rho_0^2 \omega^2} \frac{{j_n(kr)}^2}{r^2} \left( \frac{d P_n^m(\cos \theta)}{d \theta} \right)^2
\end{align}
on the $z$-axis. We can now use the following formula to compute the derivative of the associated Legendre polynomial.

\begin{align}
& \frac{d}{d \theta} P_n^m (\cos \theta) \\
& = - \frac{1}{2} \left[ (n+m)(n-m+1) P_n^{m-1} (\cos \theta) - P_n^{m+1} (\cos \theta) \right] \nonumber \\
&  \mbox{for } 1 \leq m \leq n-1 \mbox{, and }\nonumber \\
& = - \frac{1}{2} \left[ (n+m)(n-m+1) P_n^{m-1} (\cos \theta) \right] \mbox{for } m = n \nonumber
\end{align}
In this formula,  $P_n^{m+1}$ is always equal to 0 on the $z$-axis (since $(m+1) >0$), while $P_n^{m-1}$ is $\neq 0$, only for $m=1$. Thus only SBV of topological order $m = 1$ can exert an axial force. This can be seen on Fig. \ref{Figure3}b and \ref{Figure3}d. In addition, the radial evolution of the vortex is given by the function $j_n(kr)^2/r^2$ which is maximum at the center $r=0$ only for $n=1$ and minimum (zero) otherwise ($n \geq 2$). This means that only the SBV of order $(n,m) = (1,1)$ can trap a particle axially in the LWR.

\subsection{Summary of Section \ref{sec: Single SBB}}

The results of this section are summarized in tables \ref{table2} and \ref{table3}. This analysis shows that for particles with positive contrast factors $f_1$ and $f_2$, a stable lateral trap can only be produced by a SBV of order $(n,m)$ when $(n+m)$ is even. For $(n,m) = (1,1)$ both the potential and kinetic energy contribute to the trap while for $(n+m)$ even and $(n,m) \neq (1,1)$, the potential energy contributes to the trap while the kinetic energy tends to expel the particle. In addition, $(n,m) = (1,1)$ produces the strongest lateral gradients and hence lateral trapping force as can be seen on Fig. \ref{Figure3}a. Concerning the axial trap, only the kinetic energy contributes to the axial force and a trap is only obtained for $(n,m) = (1,1)$, which is the only SBV wherein the kinetic energy is maximum at the center. Hence, since only the SBV of order $(n,m) = (1,1)$ enables to obtain both a lateral and axial trap, we will mainly consider this case in the remaining part of this work. Note that the results obtained here for positive contrast factors presented in table \ref{table2} and \ref{table3} can be extended to negative contrast factors by simply inverting the ``Trap" and ``Expel" words in the tables \ref{table2} and \ref{table3}. 

\begin{table}[!htbp]
\small
  \caption{Lateral force for positive contrast factors}
  \label{table2}
  \begin{tabular}{l c c }
\hline
 & Potential energy & Kinetic energy \\
\hline 
$(n+m)$ odd  & No force & Expel \\

$(n+m$) even, $(n,m) = (1,1)$ & Trap & Trap \\ 

$(n+m)$ even, $(n,m) \neq (1,1)$  & Trap & Expel \\
 \hline
  \end{tabular}
\end{table}

\begin{table}[!htbp]
\small
  \caption{Axial force for positive contrast factors}
  \label{table3}
  \begin{tabular}{l c c}
\hline
 & Potential energy & Kinetic energy \\
\hline
$m \neq 1$ & No force & No force \\

$(n,m) = (1,1)$  & No force & Trap \\

$m=1$, $n > 1$ & No force  & Expel \\
\hline
  \end{tabular}
\end{table}

\section{\label{sec:3D assembly} 3D particle assembly with synchronised spherical Bessel vortices}

\subsection{\label{sec: III A} Synchronized spherical Bessel vortices}
\begin{figure}[!htbp]
\centering
  \includegraphics[width=7.8cm]{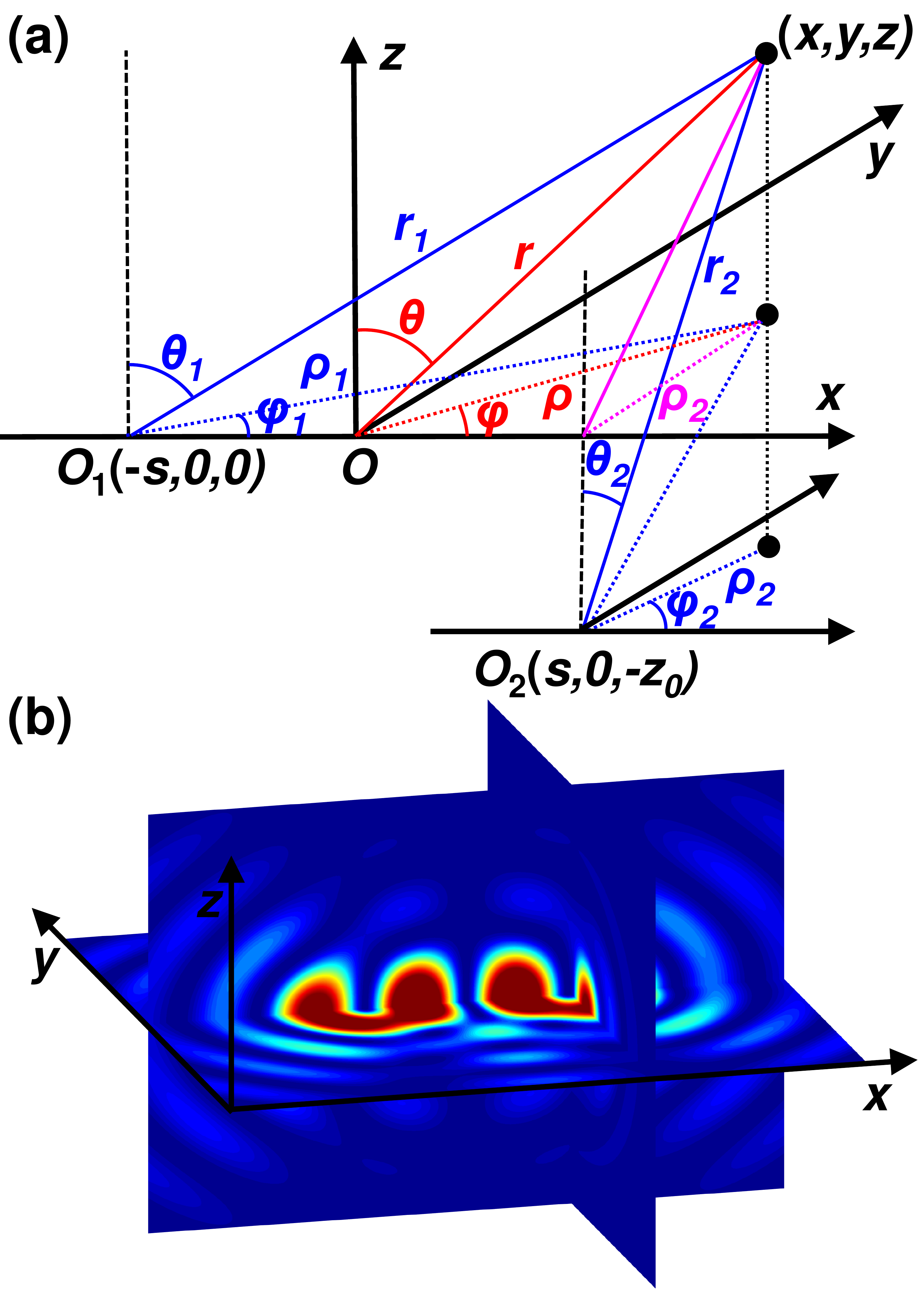}
  \caption{(a) Local and global coordinates relationship in spherical coordinates system and (b) Square of the pressure modulus $|p|^2$ of two interacting synchronized SBVs of order $(n,m)$=(1,1) with lateral offset ratio $\delta$=2 ($\delta=s / l_0 $) and no axial offset $\alpha=0$ ($\alpha=z_0 / l_0 $). 
  $l_0$ is the distance between the maximum pressure amplitude on the first ring and the origin of a single vortex beam.}
  \label{fig2:Schematic and coordinates of two spherical BBs.}
\end{figure}
The ability to trap a particle both laterally and axially in the LWR with a SBV of order $(1,1)$ has been demonstrated in the previous section. Note that three-dimensional trapping of a PS particle with a one-sided spherical vortex of order $(1,1)$ beyond the long wavelength approximation has been demonstrated both theoretically and experimentally in \cite{baresch2013spherical,baresch2016observation} while lateral trapping has been demonstrated at micrometric scales in \cite{baudoin2019folding,nc_baudoin_2020}. However, a central question is: how assembling in 3D particles with spherical vortices since the ring surrounding the trapped particle is repulsive for particle located outside (as can be seen on Fig. \ref{Figure3})? In our previous work, we demonstrated theoretically the ability to assemble in 2D particles trapped at the core of 2 cylindrical vortices by using their interference, which creates an attractive path between the two vortices. We now examine the 3D case with spherical vortices.

Consider two SBVs separated by a lateral offset $2s$ and an axial offset $z_0$, as shown in Fig. \ref{fig2:Schematic and coordinates of two spherical BBs.}a.
The acoustic pressure perturbation of each individual beam can be described by the formula:
\begin{equation}
p_{j}=A_{j} j_{n}\left(k r_{j}\right) P_n^m (cos\theta_j) e^{i m \varphi_{j}} e^{i \beta_{j}},
\label{Eq4 individual pressure}
\end{equation}
where the index $j=1$ denotes the left vortex with beam center $O_1 (-s,0,0)$ in global coordinates, $j=2$ is the lower right vortex with beam center $O_2 (s,0,-z_0)$. $\beta_{j}$ is the additional phase angle coming from the temporal term. Here we will study synchronized vortices, i.e. vortices with no phase shift: $\beta_{1} = \beta_{2}$. Inded, the case of non-synchronized vortices interaction is studied for cylindrical Bessel beams in \cite{prap_gong_2019} and showed that optimal conditions are obtained when the two vortices are perfectly synchronized. Since the origin of time can be set arbitrarily, we will thus consider $\beta_{1} = \beta_{2} = 0$ in the following calculations. The total acoustic field is the superposition the two SBVs:
\begin{equation}
\begin{aligned}
p = & A_{1} j_{n}\left(k r_{1}\right) P_{n}^{m}\left(\cos \theta_{1}\right) e^{i m \varphi_{1}}+ \\
& A_{2} j_{n}\left(k r_{2}\right) P_{n}^{m}\left(\cos \theta_{2}\right) e^{i m \varphi_{2}},
\label{Eq5 Total pressure}
\end{aligned}
\end{equation}
where the geometrical relationship of the position and angle coordinates are given in detail in Appendix \ref{sec:Appendix B}1. The pressure amplitude square $|p|^{2}$ of two synchronized interacting SBVs is illustrated in Fig. \ref{fig2:Schematic and coordinates of two spherical BBs.}b for a lateral offset ratio $\delta=s/l_0 = 2$ and an axial offset ratio  $\alpha=z_{0}/l_{0} = 0$, with $l_0$ is the distance between the maximum pressure amplitude on the first ring (corresponding to the position of the first maximum of the spherical Bessel function) and the origin of a single vortex beam. In the present simulations, $l_0= 100 \; \mu$m. The maximum pressure amplitudes of the two individual beams are set the same: $| p_{max} |$=1 MPa. The interference between the two vortices as a function of their offset can be seen on the Movie 2 in SI for different SBV orders. In order to obtain the radiation force exerted by the combination of two SBVs on small particles compared to the wavelength, we must (i) compute the acoustic velocity field associated with the total pressure field given by Eq. (\ref{Eq5 Total pressure}) with the relation: $\mathbf{v}=-i /(\rho_0 \omega) \nabla p$ with $\nabla=\partial / {\partial r} \; \mathbf{e}_{r}+ \partial / {r \partial \theta} \; \mathbf{e}_{\theta}+{\partial } / { (r \sin \theta)\partial \varphi} \; \mathbf{e}_{\varphi}$, (ii) compute Gor'kov's potential with Eq. (\ref{gorkov}) and finally (iii) calculate the negative gradient of this potential. The detailed derivation of the velocity field as well as related derivative relationships are given in Appendix \ref{sec:Appendix B}2.

\subsection{\label{sec: III B}Particle assembly along the lateral direction}
\begin{figure*} [!htbp]
\includegraphics[width=17.8cm]{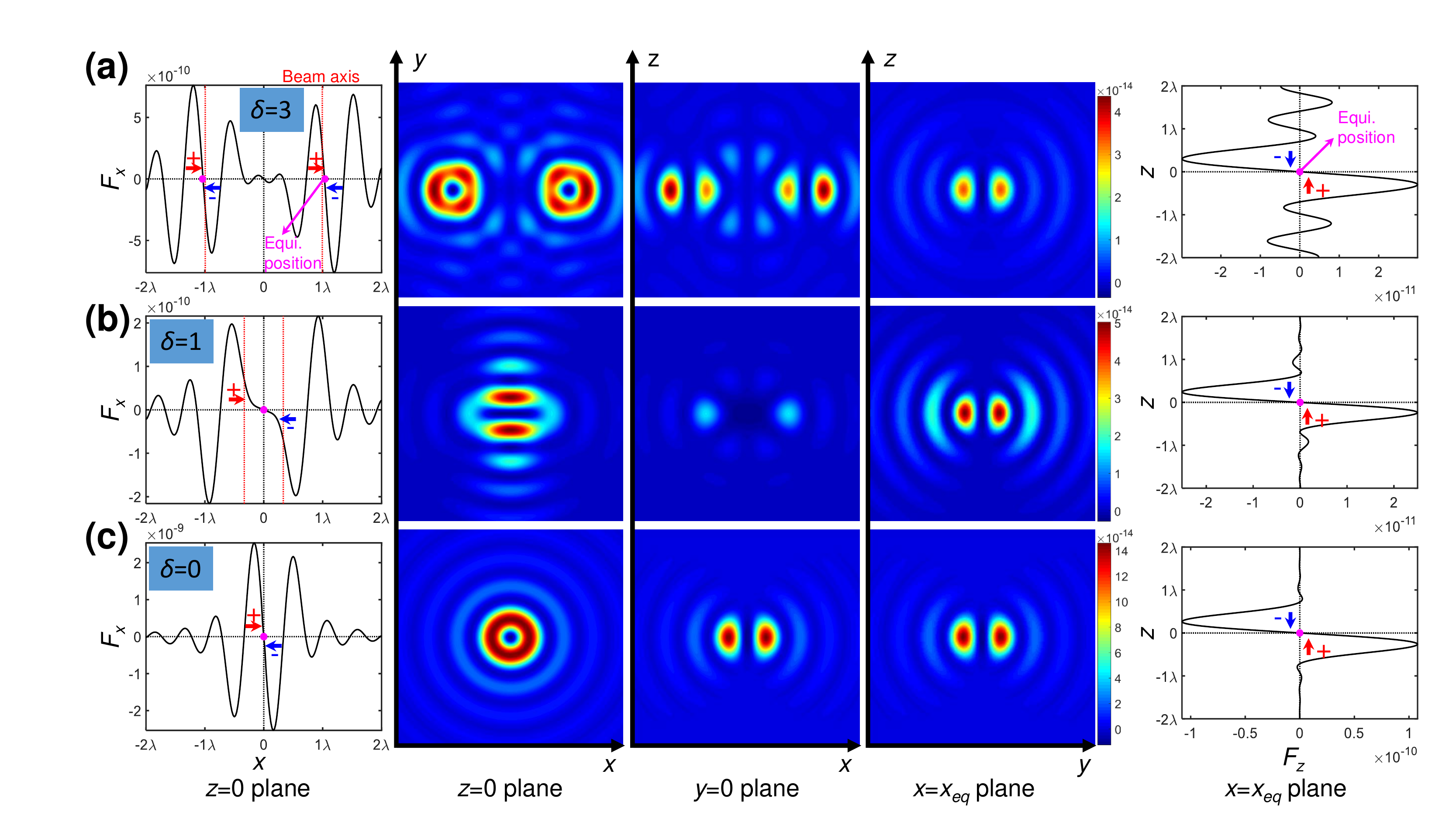}
\caption{Particles assembly of particles trapped at the center of two SBVs of order (1,1) along the lateral ($x$) direction. Columns 2 to 4 represent the Gor'kov potential in the $(x,y)$, $(x,z)$ and $(y,z)$ planes respectively. The lateral and axial radiation forces vs positions are represented on the first and fifth columns, respectively. The 3 rows (a), (b) (c) correspond to  different offset ratios $\delta$ (dimensionless distance between the vortices core): (a) $\delta$=3; (b) $\delta$=1; (c) $\delta$=0. The particles get trapped at the static equilibrium position indicated by the magenta solid spheres in lateral and axial directions with restoring forces (negative force gradient). The continuous evolution of the Gork'ov potential and the axial and lateral forces as a function of positions with different offset ratios can be seen in the Movie 3 in SI. The results demonstrate theoretically the ability to assemble laterally particles with synchronized vortices, while maintaining an axial trap.}
\label{Fig3:Particle assembly in x derection.}
\end{figure*}
First, we study the lateral assembly of two particle with two spherical Bessel beams of order (1,1) with no offset in the axial direction (i.e., $z_0 = 0$). To enable 3D assembly, the separate particles should always remain trapped in the axial direction, while being pushed laterally without the possibility for the particles to escape from the potential well. Simulations are performed for two PS particle of radius $a=5 \; \mu$m trapped at the center of two first order ($(n,m)=(1,1)$) SBVs with driving frequency $f$= 5 MHz and maximum beam amplitude $|p_{max}| = 1$ MPa.  Slices of the Gor'kov potential in the 3 planes (i) $(x,y)$, $z=0$, (ii) $(x,z)$, $y=0$ and (iii) $(y,z)$, $x=x_{eq}$ (defined below) are represented in Fig. \ref{Fig3:Particle assembly in x derection.} in the columns 2 to 4 respectively at different lateral offset ratios in the $x$ direction (along which the particles are assembled) $\delta=$ 3, 1 and 0 (rows (a) to (c)). The computational domain is $x,y,z \in [-2\lambda,2\lambda]$ with $\lambda$=301 $\mu$m in water. The axial and lateral radiation force are also represented on this figure (first and last column respectively) as well as the static equilibrium positions for the two particles pointed by magenta solid spheres. Note that the $x$ coordinate of the two equilibrium positions of the particles ($x = \pm x_{eq}$) differ from the position of each individual vortex central axis when $\delta \neq 0$. The continuous evolution of the Gor'kov's potential and the lateral and axial forces for $\delta$ evolving from 5 to 0 can be seen on the Movie 3 in SI.

As observed from the second row of Fig. \ref{Fig3:Particle assembly in x derection.} and Movie 3, the interference between the two vortices creates a path in the repulsive rings, which enables the assembly of two particles initially trapped individually at the center of the two vortices. All along the way, each of the particle is trapped axially and is pushed by a lateral force along the $x$-axis until the two particles are gathered at the core of the two superimposed vortices ( Fig. \ref{Fig3:Particle assembly in x derection.}c). This demonstrates the ability for lateral assembly while maintaining an axial trapping with two synchronized SBVs.

\subsection{\label{sec: III C} Particle assembly along the axial direction}
\begin{figure*} [!htbp]
\includegraphics[width=17.6cm]{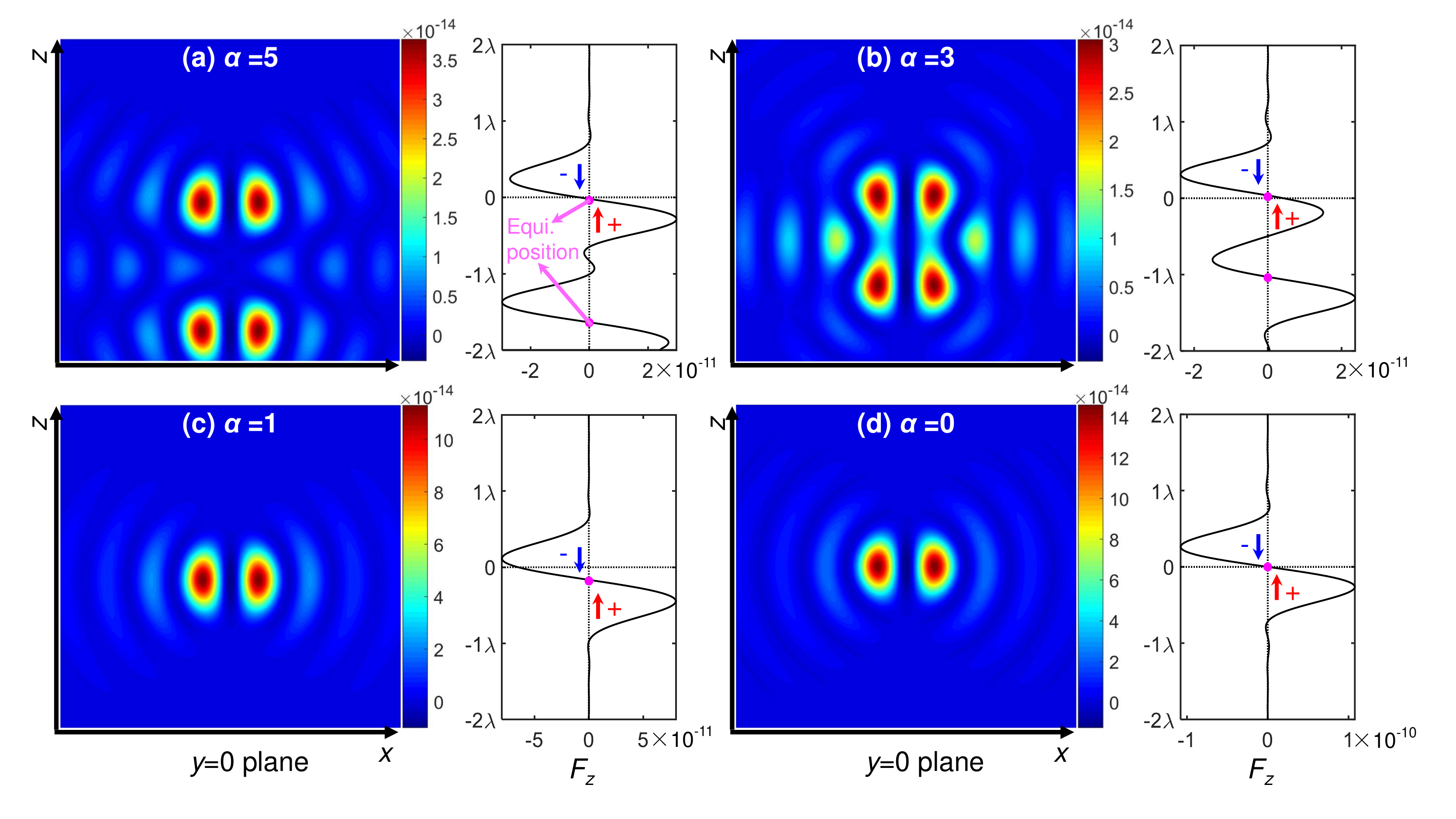}
\caption{Particles assembly of particles trapped at the center of two synchronized SBVs of order (1,1) along the axial ($z$) direction. The Gor'kov potential (color map) and axial force $F_z$ are represented for four different axial offset ratios (dimensionless distance between the vortices center) (a) $\alpha$=5, (b) $\alpha$=3, (c) $\alpha$=1, and (d) $\alpha$=0. 
The continuous evolution of the Gork'ov potential and the axial forces as a function of $z$ position can be seen in the Movie 4 in SI for $\alpha$ evolving from 5 to 0.}
\label{Fig4:Particel assembly in z direction.}
\end{figure*}

In 3D, both lateral and axial assembly are interesting to assemble complex objects. We now focus on the axial assembly. Hence we consider two vortices whose central axis coincide, while they are separated by an axial offset ratio $\alpha = z_0 / l_0$. We focus below on the axial trap, but the lateral trap remains effective all along the assembly process. The Gor'kov potential of PS particles trapped in synchronized SBV field with different axial offset ratios (a) $\alpha = 5$, (b) $\alpha = 3$, (c) $\alpha = 1$ and (d) $\alpha = 0$ are described in Fig. \ref{Fig4:Particel assembly in z direction.}. Similarly, the computational region is $x,y,z \in [-2\lambda,2\lambda]$. The Gor'kov potential is symmetric around the $x$ axis (i.e., $x=0$) in the $(x,z), y=0$ plane, as shown in the colormap panels in (a-d) of Fig. \ref{Fig4:Particel assembly in z direction.}. To verify the assembly ability along the axial direction, we calculated the axial radiation force $F_z$ as a function of $z$ position for different offsets. The continuous evolution of the Gork'ov potential and the axial radiation force as a function of $z$ position can be visualized in Movie 4. The results show that two particles are pushed axially until they are assembled, as illuminated from panels (a) to (d). This demonstrates the possibility to assemble particles axially

\section{\label{sec: Critical speed} Critical speed of moving the tweezers}
\begin{figure} [!htbp]
\includegraphics[width=8.6cm]{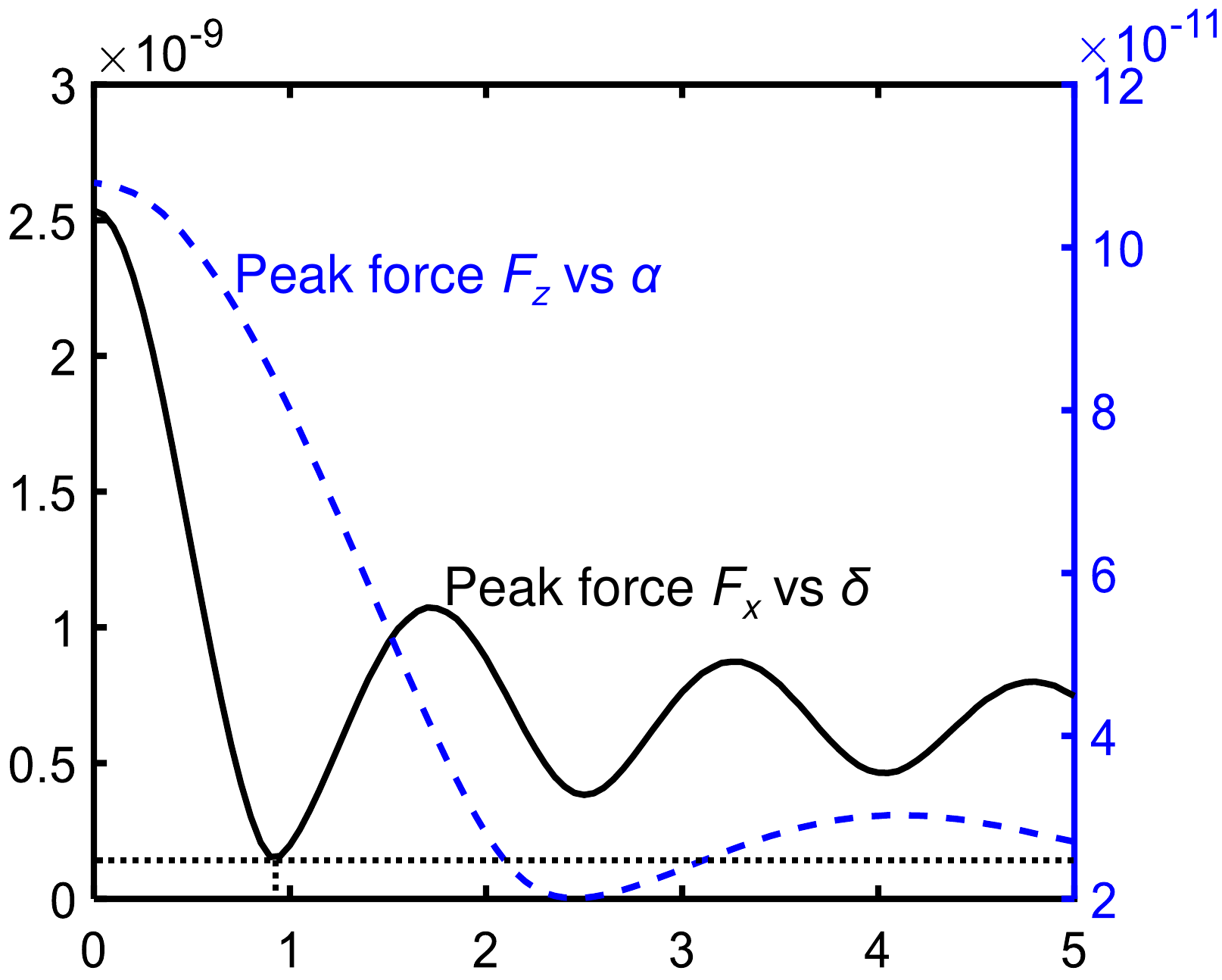}
\caption{Lateral ($F_x$) and axial ($F_z$) radiation forces (corresponding to the value of the force peak surrounding the static equilibrium position) exerted on a trapped particle in the lateral and axial assembly configurations respectively as a function of the dimensionless distance between the particles. For the lateral particles assembly, the critical radiation force $F_x^{cr}=1.538 \times 10^{-10}$ N is obtained at the lateral offset ratio $\delta$=0.9, while for the axial assembly, the critical radiation force is $F_z^{cr}=2.011 \times 10^{-11}$ N  is obtained at the axial offset ratio $\alpha$=2.5.}
\label{Fig5:Critical force in x and z direction.}
\end{figure}
The next critical question to address is the speed at which two particles can be assembled (moved). To move particles at a given speed, the acoustic radiation force in the direction of the movement must resist the Stokes' drag acting on the particle. By balancing these two forces, it is possible to determine the critical maximum speed at which ATs can be translated without losing the particle. Assuming that the particle are moved at a constant speed, the critical speed $v_{x}^{cr}$ and $v_{z}^{cr}$ in the lateral and axial direction respectively (for lateral and axial assembly) can be determined by balancing the minimum radiation force ($F_{x,z}^{cr}$) along the way (i.e. for different offsets $\delta$ or $\alpha$) and the drag force: $\mathbf{F}_{d}=- 6 \pi \eta a v_{x,z}^{cr}$, with $\eta$ the dynamic viscosity. Here the radiation force refers to the radiation force value at the peak surrounding the potential well. Its evolution in the lateral and axial assembly configurations are plotted as a function of the corresponding offset ratios ($\delta$ for the lateral assembly (black line) and $\alpha$ for the axial assembly (blue dashed line)) in Fig. \ref{Fig5:Critical force in x and z direction.}. For lateral assembly, the minimum force is obtained for $\delta=0.9$ with the critical radiation force $F_x^{cr}=1.5 \times 10^{-10}$ N. For axial assembly, the minimum lies at $\alpha=2.5$ with the critical radiation force $F_z^{cr}=2.0 \times 10^{-11}$ N. The lower force for axial assembly is expected since the axial trap is typically one order of magnitude smaller than the lateral trap for acoustical vortices. We can now compute the critical assembly speed from the relationship $F_{x,z}^{cr}= F_d$, leading to the expression $v_{x,z}^{cr}=F_{x,z}^{cr} / ( 6 \pi \eta a ) $. For a PS particle with $a=5 \mu$m immersed in water with $\eta=1.0 $ mPa s, the critical speed for lateral assembly is $v_x^{cr} = 1.6 $ mm/s, while for axial assembly, we obtain $v_z^{cr} = 0.2$ mm/s, whose values are compatible with typical operations in microfluidic systems. Note that these values of course depend on the size and composition of the particle and the intensity of the beam.

\section{\label{sec:conclusion} Conclusion and discussion}
In this work, we demonstrated theoretically the possibility to assemble in 3D (both laterally and axially) particles trapped at the center of two synchronized spherical Bessel acoustical vortices (SBVs) in the long wavelength regime (i.e. for particle much smaller than the wavelength). The assembly is enabled by the destructive interference between approaching neighboring vortices, which creates an attractive path between the trapped particles. We also determined the maximum speed at which the particles can be assembled from the balance of the radiation force and the drag force. Speeds of the order of $mm s^{-1}$ are predicted for $5 \, \mu$m polystyrene spheres, excitation frequency of $5$ MHz and reasonable pressure levels of 1 MPa. This study was performed in the analytically tractable case of SBVs trapping small particles compared to the wavelength. Nevertheless, such vortices are difficult to synthesize experimentally since transducers positioned all around the manipulation area would be required. The next step is hence to investigate theoretically and experimentally the assembly of particles with one-sided spherical vortices \cite{baudoin2020Acoustical} -- i.e. vortices synthesized with a finite aperture lower than $2 \pi$ steradian -- beyond the long wavelength approximation. Such vortices can indeed be synthesized with miniaturized, flat holographic transducers \cite{baudoin2019folding,nc_baudoin_2020}. In such studies, the secondary radiation forces \cite{doinikov2001acoustic,silva2014acoustic,wang2017sound} should be considered since it could strongly affect the particle dynamics (while this effect is less important for small-sized particles \cite{silva2014acoustic}). Finally, an important question to address in the future is the role of acoustic streaming, including bulk \cite{eckart1948vortices,prl_anhauser_2012,pre_riaud_2014,prl_baresch_2018} and boundary streaming \cite{rayleigh1884circulation,karlsen2018acoustic}. Acoustic streaming could indeed affect the particles and lead to their ejection from the trap, depending on the particle size and composition, the actuation frequency and the beam topology. Finally, the combination of more than two Bessel beams  could be promising to produce desired acoustic traps of special shape with controllable orbital angular momentum \cite{kovalev2015orbital}.
\begin{acknowledgments}
We acknowledge the support of the programs ERC Generator and Prematuration funded by ISITE Universit\'{e} Lille Nord-Europe (I-SITE ULNE). The authors would like to thank Dr. Udita Ghosh for her careful reading of the manuscript.
\end{acknowledgments}

\appendix

\section{\label{sec:Appendix A} Acoustic power required for the synthesis of an acoustic spherical Bessel Vortex of order (n,m)}

A Bessel spherical vortex can be seen as the interference between an outgoing Hankel spherical vortex of the first kind and a converging Hankel spherical of the second kind:
\begin{align}
p & =A_0 j_{n}\left(k r\right) P_n^m (cos\theta) e^{i (m \varphi - \omega t)}  \\
& = \frac{A_0}{2} [h_n^{{(1)}}(kr) + h_n^{(2)} (kr)]  P_n^m (cos\theta) e^{i (m \varphi - \omega t)} \nonumber
\end{align}
Of course since a spherical Bessel vortex is steady over $\mathbf{e}_r$, the radial intensity (Poynting) vector $\mathbf{I}$ cancels in this direction. Hence to compute the power required to synthesize a Bessel vortex, only the outgoing or converging wave must be considered. This power can hence be computed by integrating this intensity vector over a sphere of radius $r$:
\begin{eqnarray}
\mathcal{\left< \mathcal{P} \right>} & = & \iint \left< \mathbf{I} \right>. \mathbf{n} dS \\
& = & \int_{\theta = 0}^\pi \int_{\varphi = 0}^{2 \pi} \left< \mathbf{I}. \mathbf{e}_r \right> r^2 \sin \theta d\theta \, d\varphi \label{eq:int}
\end{eqnarray}
with:
\begin{align}
& \left<\mathbf{I}\right> . \mathbf{e}_r  = \mathrm{Re} \left[ p^* (\mathbf{v}.\mathbf{e}_r) \right] \\
& \mbox{and }p = A_0 h_n^{{(1)}}(kr)  P_n^m (cos\theta) e^{i (m \varphi - \omega t)}
\end{align}
This integration can be performed in the far field wherein $h_n^{{(1)}}(kr) \approx i^{-(n+1)} e^{i kr} / kr$. First we obtain:
\begin{equation}
\mathrm{Re} \left[ p^* (\mathbf{v}.\mathbf{e}_r) \right] = \frac{A_0^2 {P_n^m}^2(\cos \theta)}{\rho_0 \omega r^2},
\end{equation}
whose integration on a sphere in the far field gives:
\begin{equation}
\mathcal{\left< \mathcal{P} \right>}  = \frac{2 \pi A_0^2}{\rho_0 \omega} \times \frac{2(n+m)!}{(2n+1)(n-m)!}.
\end{equation}

\section{\label{sec:Appendix B} Coordinates relationship and synthetic velocities}

\subsection{\label{sec:Appendix B1} Geometrical relationship of local and global coordinates}
Take Fig. \ref{fig2:Schematic and coordinates of two spherical BBs.}a as a general case with offsets in lateral (here $x$) and axial ($z$) directions. The local (position and angle) coordinates are depicted in terms of global spherical coordinates as follows:
\begin{align}
\begin{split}
& \rho_{1}^{2} = \rho^{2}+s^{2}+2 \rho s \cos \varphi , \\
& \rho_{2}^{2} = \rho^{2}+s^{2}-2 \rho s \cos \varphi , \\
& r_{1}^{2} = \rho_{1}^{2}+z^{2}=r^{2}+s^{2}+2 r  s  \sin \theta  \cos \varphi , \\
& r_{2}^{2} = r^{2}+s^{2}-2 r s \sin \theta \cos \varphi+2 r \cos \theta z_{0}+z_{0}^{2} , \\
& \cos \theta_{1} = z / r_{1} = r \cos \theta / r_{1}, \\
& \cos \theta_{2} = \left(z+z_{0}\right) / r_{2} =( r \cos \theta + z_0) / r_{2} .
\label{EqApp Geometrical relationship}
\end{split}
\end{align}
Note that there is no offset in axial direction when $z_0=0$. 

\subsection{\label{sec:Appendix B2} Velocity of synchronized field}
The velocity of synchronized field could be computed by the vector sum of individual velocities as shown in Eq. (\ref{EqApp v1,2}) with the pressure expression given in Eq. (\ref{Eq4 individual pressure})
\begin{equation}
\begin{aligned}
\mathbf{v}_{1,2} &= -i \frac{1}{\rho_0 \omega} \nabla p_{1,2}  \\
& = -i \frac{1}{\rho_0 \omega}\left\{\frac{\partial p_{1,2}}{\partial r} \mathbf{e}_{r}+\frac{1}{r} \frac{\partial p_{1,2}}{\partial \theta} \mathbf{e}_{\theta}+\frac{1}{r \sin \theta} \frac{\partial p_{1,2}}{\partial \varphi} \mathbf{e}_{\varphi}\right\},
\label{EqApp v1,2}
\end{aligned}
\end{equation}

The total velocity is
\begin{equation}
\mathbf{v}=\mathbf{v}_{1}+\mathbf{v}_{2}=v_{r} \mathbf{e}_{\mathrm{r}}+v_{\theta} \mathbf{e}_{\theta}+v_{\varphi} \mathbf{e}_{\varphi},
\label{EqApp v=v1+v2}
\end{equation}
And the time average of the square of velocity is
\begin{equation}
\begin{aligned}
\left\langle\mathbf{v}^{2}\right\rangle &= \left\langle\left[v_{r} \mathbf{e}_r+v_{\theta} \mathbf{e}_{\theta}+v_{\varphi} \mathbf{e}_{\varphi}\right]^{2}\right\rangle \\
& =\frac{1}{2} \operatorname{Re}\left\{v_{r} v_{r}^{*}+v_{\theta} v_{\theta}^{*}+v_{\varphi} v_{\varphi}^{*}\right\}
\label{EqApp average v square}
\end{aligned}
\end{equation}

To compute the three components of the velocity, the following geometrical and derivative relationships should be applied

\begin{align}
\begin{split}
& \rho_{1,2} e^{i \varphi_{1,2}} =\rho e^{i \varphi} \pm s, \\
& {\partial r_{1}} / {\partial r} = ( r+ s \sin \theta \cos \varphi ) / r_1, \\
& {\partial r_{1}} / {\partial \theta} =  r s \cos \theta \cos \varphi / r_1, \\
& {\partial r_{1}} / {\partial \varphi} = - r s \sin \theta \sin \varphi / r_1, \\
& {\partial \rho_{1}} / {\partial r} = ( r \sin ^{2} \theta+ s \sin \theta \cos \varphi) / \rho_1, \\
& {\partial \rho_{1}} / {\partial \theta} = ( r^{2} \sin \theta \cos \theta+ r s \cos \theta \cos \varphi) / \rho_1, \\
& {\partial \rho_{1}} / {\partial \varphi} = - r s \sin \theta \sin \varphi / \rho_1, \\
& {\partial r_{2}} / {\partial r} = ( r- s \sin \theta \cos \varphi + 2 z_0 \cos \theta) / r_2, \\
& {\partial r_{2}} / {\partial \theta} = - r( s \cos \theta \cos \varphi +  z_0 \sin \theta)/ r_2, \\
& {\partial r_{2}} / {\partial \varphi} = r s \sin \theta \sin \varphi / r_2, \\
& {\partial \rho_{2}} / {\partial r} = ( r \sin ^{2} \theta - s \sin \theta \cos \varphi) / \rho_2, \\
& {\partial \rho_{2}} / {\partial \theta} = ( r^{2} \sin \theta \cos \theta - r s \cos \theta \cos \varphi) / \rho_2, \\
& {\partial \rho_{2}} / {\partial \varphi} = r s \sin \theta \sin \varphi / \rho_2 .
\label{EqApp Derivate}
\end{split}
\end{align}
\renewcommand\refname{Reference}
\bibliographystyle{unsrt}  
\bibliography{main}        

\end{document}